\documentclass[journal=nalefd,manuscript=article]{achemso}

\usepackage[version=4]{mhchem} 
\usepackage{graphicx}
\usepackage{xcolor}
\usepackage{comment}
\usepackage[normalem]{ulem}
\usepackage{hyperref}
\hypersetup{
    colorlinks=true,
    citecolor=blue,
    filecolor=blue,
    linkcolor=blue,
    urlcolor=blue
}
\usepackage{amsmath,amssymb}
\usepackage{pdfpages}

\usepackage{xr-hyper}
\usepackage{cleveref}

\newcommand{\didv}{${\rm d}I/{\rm d}V$}

\author{Manuel Siegl}
\affiliation{London Centre for Nanotechnology, University College London, WC1H 0AH, London, UK}
\alsoaffiliation{Department of Physics and Astronomy, University College London, WC1E 6BT, London, UK}

\author{Julian Zanon}
\affiliation{Department of Applied Physics and Science Education,
Eindhoven University of Technology, Eindhoven 5612 AZ, The Netherlands}

\author{Joseph Sink}
\affiliation{Department of Physics and Astronomy, University of Iowa, Iowa City, Iowa 52242, USA}

\author{Adonai Rodrigues da Cruz}
\affiliation{Department of Physics and Astronomy, University of Iowa, Iowa City, Iowa 52242, USA}

\author{Holly Hedgeland}
\affiliation{London Centre for Nanotechnology, University College London, WC1H 0AH, London, UK}

\author{Neil J. Curson}
\affiliation{London Centre for Nanotechnology, University College London, WC1H 0AH, London, UK}
\alsoaffiliation{Department of Electronic and Electrical Engineering, University College London, WC1E 6BT, London, UK}

\author{Michael E. Flatté}
\affiliation{Department of Physics and Astronomy, University of Iowa, Iowa City, Iowa 52242, USA}
\alsoaffiliation{Department of Applied Physics and Science Education,
Eindhoven University of Technology, Eindhoven 5612 AZ, The Netherlands}
\email{michaelflatte@quantumsci.net}

\author{Steven R. Schofield}
\affiliation{London Centre for Nanotechnology, University College London, WC1H 0AH, London, UK}
\alsoaffiliation{Department of Physics and Astronomy, University College London, WC1E 6BT, London, UK}
\email{s.schofield@ucl.ac.uk}

\title{Imaging the Acceptor Wave Function Anisotropy in Silicon}

\begin{document}

\begin{abstract}
We present the first scanning tunneling microscopy (STM) images of hydrogenic acceptor wave functions in silicon. These acceptor states appear as square ring–like features in STM images and originate from near-surface defects introduced by high-energy bismuth implantation into a silicon (001) wafer. Scanning tunneling spectroscopy confirms the formation of a $p$-type surface. Effective-mass and tight-binding calculations provide an excellent description of the observed square ring–like features, confirming their acceptor character and attributing their symmetry to the light- and heavy-hole band degeneracy in silicon. Detailed understanding of the energetic and spatial properties of acceptor wave functions in silicon is essential for engineering large-scale acceptor-based quantum devices.
\end{abstract}

Shallow defects associated with substitutional impurities in semiconductors are known to be stable and reproducible localized quantum systems. These shallow states, with binding energies of a few tens of meV in silicon~\cite{YuCardona}, are well described using a hydrogen-like wave function that extends over many atoms, couples to electric and magnetic fields, and provides access to both charge and spin degrees of freedom. The spin states associated with these defects provide an exceptionally stable platform for quantum technologies, especially quantum computing. When housed in the pristine environment of isotopically purified and low defect density silicon, these spin states exhibit notably long coherence times and high resistance to noise~\cite{Muhonen2014,Kobayashi2020}. These properties, combined with the fact that each qubit is naturally identical, position dopants in semiconductors as an arguably superior choice compared to other physical qubit implementations~\cite{Weber2010,Koenraad2011}.

In silicon, nanometer positioning of individual phosphorus~\cite{Schofield2003}, arsenic~\cite{Stock2020,Stock2024}, and boron~\cite{Skeren2020} donor and acceptor impurities has been achieved via scanning tunneling microscopy (STM). This has led to the fabrication of remarkable atomic-scale electronic devices, including the single-atom transistor~\cite{Fuechsle2012} and few-donor qubit gates~\cite{He2019,Thorvaldson2024}. In order to fully utilize these devices in technologies, such as in the qubit array of an error corrected quantum computer~\cite{Hill2015a}, the fabrication process will need to be scaled to produce millions of deterministically placed dopants with predetermined and/or gate tunable couplings. This will require not only exquisite control over the positioning of these dopants, but also a detailed understanding of the spatial properties of single defects and their interactions with each other and their environment.

In compound semiconductors, like GaAs, there is a long history of dopant investigations using cross-sectional STM where the semiconductor is cleaved in situ~\cite{Ebert2003}. Notable successes include mapping the spatial structure of manganese~\cite{PhysRevLett.92.047201,Yakunin2004a,Marczinowski2007} and zinc acceptors~\cite{Loth2006}, observing binding energy shifts as a function of dopant depth~\cite{Garleff2010}, and manipulating the charge state of individual donors~\cite{Teichmann2008}. However, in silicon this cleavage method works well only for the (111) surface whose surface states have hindered attempts to image subsurface dopant wave functions~\cite{Studer2012}. 

Efforts to image dopant wave functions in silicon have focussed on thermally-prepared silicon (001) surfaces. Early measurements reported imaging the screened Coulomb potential of ionized arsenic and boron defects at room temperature~\cite{Liu2001,liu2002a,Nishizawa2008}. Later, measurements of phosphorus and boron dopants at 4.2~K reported tunneling through neutral donor and acceptor states, respectively~\cite{Miwa2013,Mol2013}. In both cases, no spatial structure of the hydrogenic state was observed, possibly due to the shallow depths of the dopants beneath the surface. Nevertheless, subsequent experiments observed highly anisotropic localized features attributed to the hydrogenic donor wave functions of the substitutional arsenic~\cite{Sinthiptharakoon2014,Usman2016}. Within the top 12 layers ($\sim1.5$~nm) of the surface, the observed features are strongly influenced by the $2\times1$ reconstruction of the hydrogen-terminated silicon (001) surface~\cite{Brazdova2015}. For arsenic donors deeper beneath the surface ($\sim2.5$~nm), the observed features consist of a 1s type envelope function modulated by a high frequency Bloch component observable due to the effect of valley interference in the silicon conduction band~\cite{Salfi2014,Voisin2020}. 

\begin{figure}
\centering
\includegraphics[width=8.5cm]{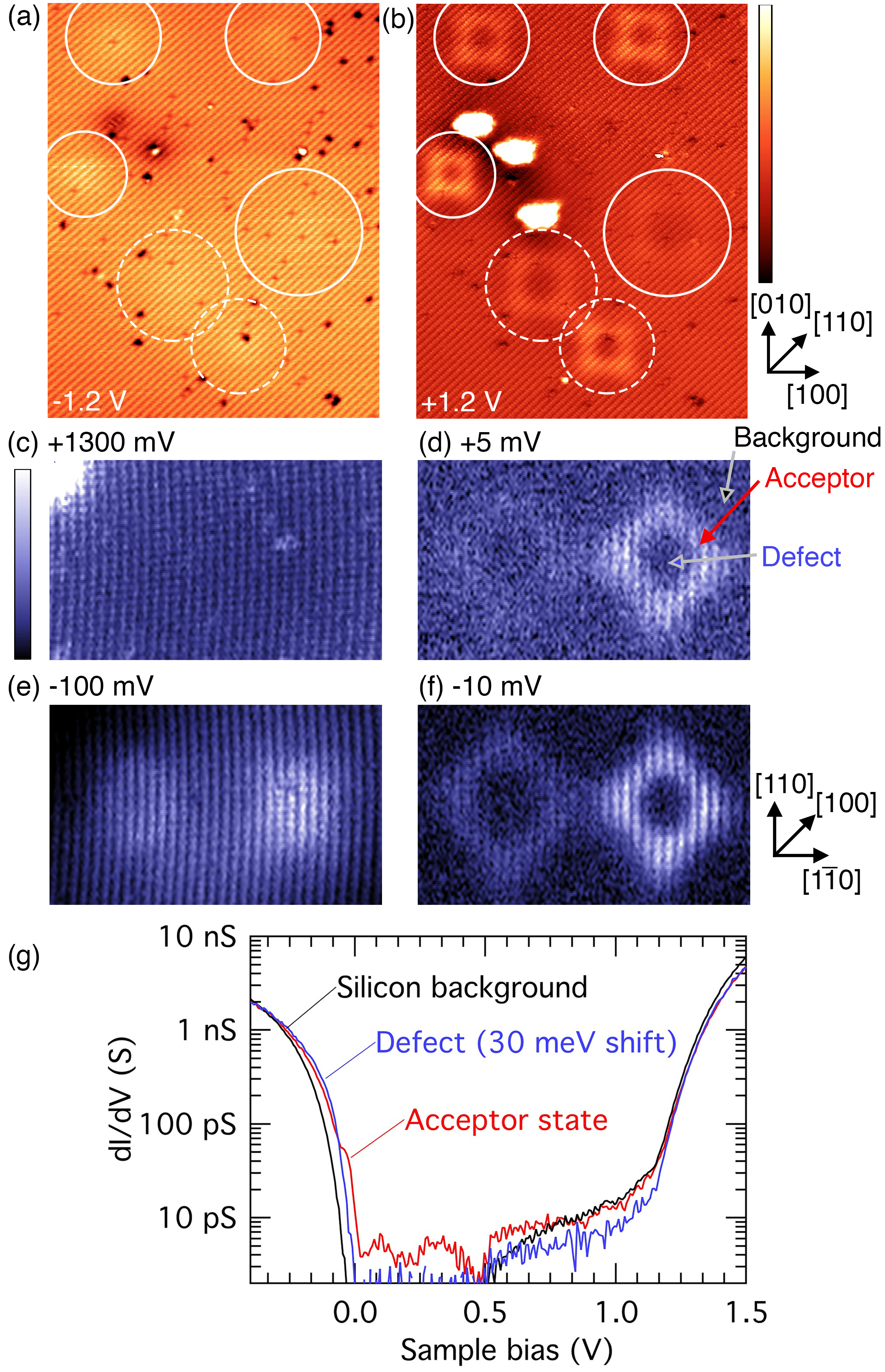}
\caption{STM images of hydrogenic acceptor states beneath a Si(001) surface. (a,b) Filled- and empty-state images of the same $54$~nm$~\times 38$~nm region of the surface. The locations of six acceptor states are highlighted by circles.  Small black features are surface vacancy defects in the surface layer. (c-f) Differential conductance (\didv) maps over the two acceptors  indicated by dashed circles in {\protect Fig.~\ref{fig1}a,b} at bias voltages $+1300$, $+5$, $-100$ and $-10$~meV, respectively. (g) Spatially integrated tunneling spectroscopy showing: (black) the silicon background away from the acceptor states; (red) the full region inside the square acceptor state of the rightmost acceptor; (blue) the central depression region (only) of the rightmost acceptor. The regulation parameters during the \didv\ measurement were $-500$~mV and 500~pA.}
\label{fig1}
\end{figure}

Here we present the first STM observations of hydrogenic \emph{acceptor} state wave functions in silicon. The observed features are large, sometimes exceeding 10~nm in lateral extent, and are highly anisotropic, exhibiting a square-shaped enhancement with edges aligned to the $\langle001\rangle$ directions, and a central depression, giving an appearance that we describe as a ``square ring'' (Fig.~\ref{fig1}b,d,f). Using the effective‑mass approximation (EFMA) and atomistic tight‑binding (TB) calculations, we show that the observed anisotropy is consistent with its origin in the ground‑state multiplet of a hydrogenic acceptor in silicon. Our theory takes into account both the degeneracy of the light and heavy hole bands at the $\Gamma$-point as well as the local tetrahedral symmetry of the host crystal lattice~\cite{Yakunin2004a}. A slightly larger spatial extent of the observed features is accounted for in the EFMA theory by a small change in the effective mass of the (assumed isotropic) heavy hole band, which we attribute to strain in the near-surface region. The best fits for our EFMA and TB calculations are obtained for different acceptor depths.  This discrepancy is due to differing symmetry and accuracy of the two band structure models and impurity potential models.

Acceptor states were created by ion implanting a 15~$\Omega$cm arsenic-doped Si(001) wafer using five implantation energies (0.25--2.0~MeV) and corresponding fluences (Supporting Table~S1) to produce a bismuth density of $\sim1\times10^{20}\,\text{cm}^{-3}$ extending from 20 to 550~nm into the sample bulk (Supporting Fig.~S1). The sample was annealed at 500$^\circ$C overnight under ultrahigh vacuum ($<2\times10^{-10}$~mbar), followed by a 10~s flash anneal at 1150$^\circ$C by direct current heating, which repairs implantation damage and produces an atomically flat surface suitable for STM imaging.  This bismuth implantation and annealing protocol is known to produce a high concentration of near-surface acceptor centers that persist even after thermal treatment, resulting in a $p$-type surface atop a deeper $n$-type region~\cite{Peach2018}. 

Prior to STM imaging, the surface was passivated with hydrogen by being exposed to an atomic hydrogen beam (chamber background $5\times10^{-7}$ mbar for 5 minutes) while the sample was maintained at 340$^\circ$C.  STM images were taken at 77~K. Filled and empty-state STM images of the same region of an implanted sample are shown in Fig.~\ref{fig1}a,b and we have highlighted the location of six acceptor states. In the filled-state image (Fig.~\ref{fig1}a), the acceptors are barely noticeable, producing only slight isotropic enhancement superimposed on top of the silicon dimer rows that can be seen running diagonally across the image. In contrast, the empty-state image (Fig.~\ref{fig1}b) shows very striking long-range features that are square shaped with edges, sometimes longer than 10~nm, aligned along the $\left<100\right>$ directions, and a central depression (a square ring). While each of the six acceptors has the same qualitative appearance, the width and the relative intensity of the features compared to the atomic corrugation vary from feature to feature. 

Fig.~\ref{fig1}c-f presents spatially resolved conductance (\didv) measurements taken at the site of the two acceptors highlighted by dashed circles in Figs.~\ref{fig1}a,b (measurements at other acceptor sites exhibited similar behavior). At high positive bias ($+1.3$~V; Fig.~\ref{fig1}c) we resolve only the surface atomic corrugation in the \didv\ map. This indicates an absence of any defect-induced band bending, demonstrating that the acceptors are electrically neutral and that the conduction band states dominate the tunneling. At low positive and negative bias voltages ($+5$, $-10$~mV; Fig.~\ref{fig1}d,f), the acceptors appear as square ring-like features. At lower negative voltage ($-100$~mV; Fig.~\ref{fig1}e), the acceptors appear as isotropic round elevations superimposed on the atomic lattice, indicating the band-bending on the valence band states due to the negative charge of the ionized acceptors. This bias-dependent behavior is well-known from STM/STS investigations of acceptors in GaAs~\cite{Yakunin2004a}; when imaging the filled-states, the tip-induced band bending pulls the acceptor level below the Fermi level causing it to donate its hole to the valence band; conversely when imaging the unoccupied-states, the acceptor level is above the Fermi level and the hole is captured at the acceptor site and the defect becomes charge neutral~\cite{Yakunin2004a,Teichmann2008}.  The two \didv\ maps at $+5$ and $-10$~meV (Figs.~\ref{fig1}d,f) are within the substrate band gap, and therefore we do not image the silicon dimer row background surrounding the defects. Instead, we obtain tunneling current only when directly probing the acceptor state. The acceptor state is energetically at the Fermi level, and therefore we are able to inject current into the acceptor state at low empty state bias (Fig.~\ref{fig1}f) and extract current through the acceptor state at low filled state bias (Fig.~\ref{fig1}d).

Fig.~\ref{fig1}g shows tunneling spectra as a function of tip bias, generated by spatially integrating the \didv\ data (i.e., the data from which the maps shown in Fig.~\ref{fig1}c-f are sampled) over different spatial regions as indicated by arrows on Fig.~\ref{fig1}d. The black trace corresponds to the silicon background away from the acceptor states; the Fermi level lies at the valence band edge, indicating a strongly $p$-type surface, consistent with Hall effect measurements of samples prepared with a similar bismuth implantation procedure~\cite{Peach2018}. Comparing this to the spectrum within the defect depression (blue trace), we observe a 30~meV upward shift of the valence band edge, confirming the defect’s negative charge state, as expected for an ionized acceptor. Extending the integration area to include the full square-shaped enhancement (red trace) reveals a pronounced shoulder near zero bias, which is the spatially resolved acceptor state seen in Figs.~\ref{fig1}d,f.

\begin{figure}
\centering
\includegraphics[width=8.5cm]{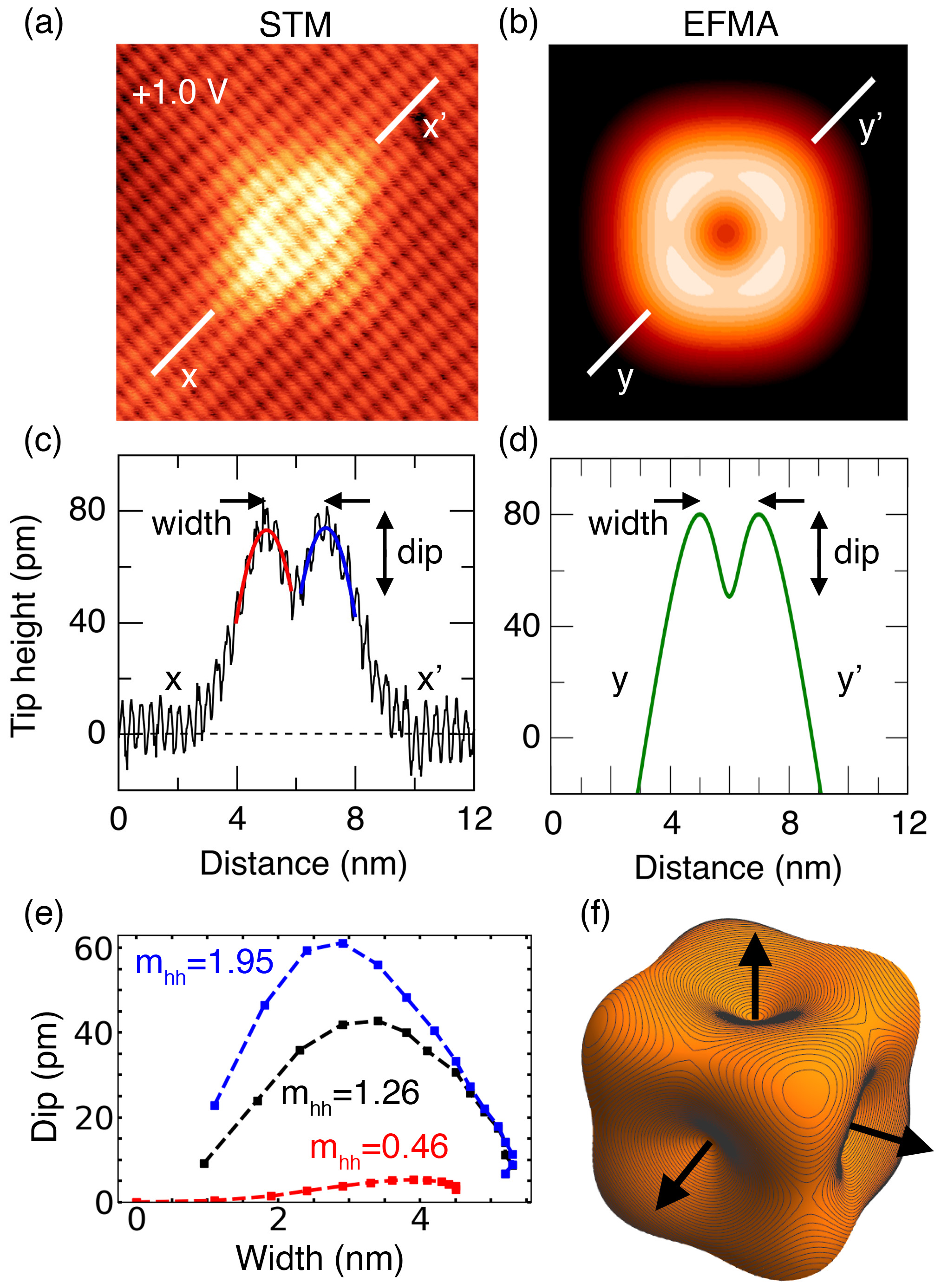}
\caption{Comparison between STM measurements and EFMA theory calculations of acceptor states in silicon. (a) STM topograph acquired at +1.0~V, 20~pA (z-range 160~pm) showing a square ring-like feature; the white line indicates the line profile shown in (c). (b) EFMA density plot of the defect wave function, calculated for an acceptor 2.9~nm below the image plane along the [001] direction. Color range corresponds to 80~pm. (d) Line profile extracted from the effective mass model in (b) for direct comparison to the experimental data in (c). For the purpose of quantifying the features, we define the parameters ``width'' and ``dip'' as shown in panels (c,d). Both experiment and theory exhibit a dip of approximately 30~pm and a width of 2.0~nm. (e) Simulated dip versus width curves for different heavy-hole effective masses ($m_{hh}$), with each point corresponding to a (001) plane spaced by 1 nm between 1.5 nm and 15.5 nm depth. (f) Isosurface of the probability density ($|\Psi|^2$) at $10^{-3},\text{nm}^{-3}$, generated by scaling the probability density by $e^{qr}$, where $q$ is analogous to an inverse Bohr radius (see Supporting Information). The results in (b) and (f) were obtained using $m_{lh} = 0.16$, $m_{hh} = 1.261$ ($\beta = 0.127$), and $\kappa = 1.15,\text{\AA}^{-1}$.}
\label{fig2}
\end{figure}

By measuring line profiles along the [110] direction, we determine the relationship between the intensity and width of the acceptor states as shown in Fig.~\ref{fig2}a,c. A detailed analysis of 75 separate acceptors, shown in Supporting Fig.~S2b,c, shows the intensity of the acceptor states, measured relative to the surrounding surface plane, decays exponentially with the feature width. This behavior correlates well with the expectation that we probe acceptors across a range of depths beneath the surface.  As the depth of the acceptor increases, the overlap of its wave function with the wave function of the STM tip decreases exponentially, while at the same time the lateral distance from the center of the defect to the region of maximum probability density increases. We have also imaged the acceptor states as a function of bias, and find the only change is a decrease in their relative intensity compared to the background (Supporting Fig.~S2a). This is expected due to an increasing contribution to the tunneling current from the bulk bands. From this observation, we can rule out contributions to the appearance of the state from quasiparticle interference or Friedel oscillations. 

The square-ring wave function observed in STM/STS experiments and described above in Figs.~\ref{fig1} and \ref{fig2}a,c is indicative of multiple Bloch states being involved in the description of the defect state, necessitating the use of multi-band theories such as EFMA and TB Green's functions. These methods have been successfully applied in past works \cite{Yakunin2004a, Yakunin2007} to describe STM features of Mn acceptors  in GaAs.

The effective mass modeling presented here used the Luttinger Hamiltonian within the spherical approximation, with the impurity described by a zero-range potential. This model was first solved analytically in Ref.~\citenum{averkiev1994spin} and later those solutions were used to construct the cubic symmetric ground state for an acceptor produced by a single Mn impurity in GaAs \cite{Yakunin2004a}. As explained in the Supporting Information (Section~2.1), our calculations obtain the cubic-symmetric solutions for the four-fold symmetric ground state ($J=3/2$) multiplet of the acceptor. These four solutions are described by the (degenerate) binding energy $E_{0}$, and four wave functions $\Psi_{M}^{3/2} = \Psi_{M}^{3/2}(r, \theta, \varphi)$, where $M$ labels $\pm 3/2$, $\pm 1.2$. The four-fold degeneracy of the multiplet will produce a measurement of the wave function that is a temporal average of the four $M$ states. As time-reversal symmetry requires the two light-hole states ($\pm 1/2$) to have identical local densities of states, and likewise the two heavy-hole states ($\pm 3/2$), we expect 
\begin{equation}
\big|\Psi_{\rm {avg}}\big|^{2} = \frac{1}{2}\big( |\Psi_{1/2}^{3/2}|^{2}+
\Psi_{3/2}^{3/2}|^{2} \big)\,. \label{eq: first_avg}
\end{equation}
Assuming a spherically symmetric delta-function like potential, centered at $r=0$ and the averaging described in Eq. \ref{eq: first_avg}, the analytic solution for the defect wave function is,
\begin{align}
\big|\Psi_{\text{avg}}(r, \theta, \varphi)\big|^{2} =\frac{\big|R_{0}(r)\big|^{2} }{16\pi}+5f(\theta, \varphi)\frac{\big|R_{2}(r)\big|^{2}}{16\pi} ,\label{eq:eq_1_eta0_final}
\end{align}
where $R_{0}(r)$ and $R_{2}(r)$ are radial functions that depend on the binding energy $E_0$ and the mass ratio $\beta = m_{lh}/m_{hh}$, where $m_{lh}$ and $m_{hh}$ are the light- and heavy-hole effective masses at the $\Gamma$ point. The presence of the angular function \( f(\theta, \varphi) \) in the second term of Eq.~(\ref{eq:eq_1_eta0_final}) causes \( \left|\Psi_{\text{avg}}(r, \theta, \varphi)\right|^2 \) to exhibit a $d$-like orbital character.

To make comparisons with the STM measurements, it is useful to consider the relationship between the tunneling current, the local density of states, and the defect wave function. The experimental STM images were collected in constant current mode, where the tip height is continuously adjusted to maintain a fixed tunneling current while scanning over the sample surface. The tunneling current reflects the probability of an electron transferring between the STM tip and the sample through vacuum, and depends sensitively on both the local density of states and the tip-sample separation. In the simplest case, assuming a delta-function tip with a constant density of states, the tunneling current is given by
\begin{align}
    I(\omega, r, \theta, z) \sim \eta(\omega, r, \theta, 0)\, e^{-2\kappa z},
\end{align}
where \( z \) is the tip-sample separation, \( \omega \) is the energy corresponding to the applied bias, and \( \kappa \) is the decay constant of the tunneling current. When on resonance with the defect, the sample local density of states $\eta$ is dominated by the defect and is well approximated by
$\eta=|\psi|^2$. Solving for the STM height expressed as a function of the defect probability density yields
\begin{equation}
    \rm{Height}(r, \theta, \varphi) = \frac{1}{2\kappa}\rm{ln}\Big(\frac{\big|\Psi_{\text{avg}}(r, \theta, \varphi)\big|^{2}}{\big|\Psi_{\text{avg}}(r_{0}, \theta_{0}, \varphi_{0})\big|^{2}}\Big),\label{eq:eq_height}
\end{equation}
where we have included an offset to set the maximum of the wave function located at $\mathbf{r}_0$ corresponding to $\rm{Height}=0$.

In Fig.~\ref{fig2}b,d-f, we show an EFMA wave function calculated considering the impurity site 2.9 nm below the (001) cleaved plane, using $E_{0} = 30$ meV and $m_{lh} = 0.16$ and $m_{hh} = 1.261$ ($\beta = 0.127$).  This produces excellent agreement between our theoretical results and the experimentally observed feature shown in Fig.~\ref{fig2}(c,d), having a peak-peak width of $2$~nm and a dip =$30$~pm. Our effective mass model in Fig.~\ref{fig2}(f) presents clearly a similar symmetry that was reported in Ref. \citenum{PhysRevX.3.011019} for an excited state  of an acceptor in Si; here, however, we show this structure emerges for the ground state.

The optimal values for $\beta$ and the [001] plane depth were obtained using Fig.~\ref{fig2}e. For $\kappa = 1.15$\AA$^{-1}$ and the light-hole mass fixed at its bulk value ($m_{lh}=0.16$), Fig.~\ref{fig2}e shows that increasing the heavy-hole mass $m_{hh}$ enhances both the feature dip and lateral width of the simulated line profile. The deviation of $m_{hh}$ from its bulk value likely reflects the influence of the surface strain field on the hole dispersion; strain-induced splitting of bulk valence bands has a well-understood effect on valence band dispersion and anisotropy\cite{YuCardona}. In addition to this, it is well known that among tetrahedral semiconductors, Si has a substantial anisotropy for the hole masses (larger, for example, than gallium arsenide)\cite{Baldereschi1974}, therefore the optimal $\beta = 0.127$ value is an isotropic value that approximates this anisotropy. A more detailed discussion of the parameter choices, including the hole masses and tunneling decay constant $\kappa$, is provided in the Supporting Information (Section~2.1).

\begin{figure}
\centering
\includegraphics[width=8.5cm]{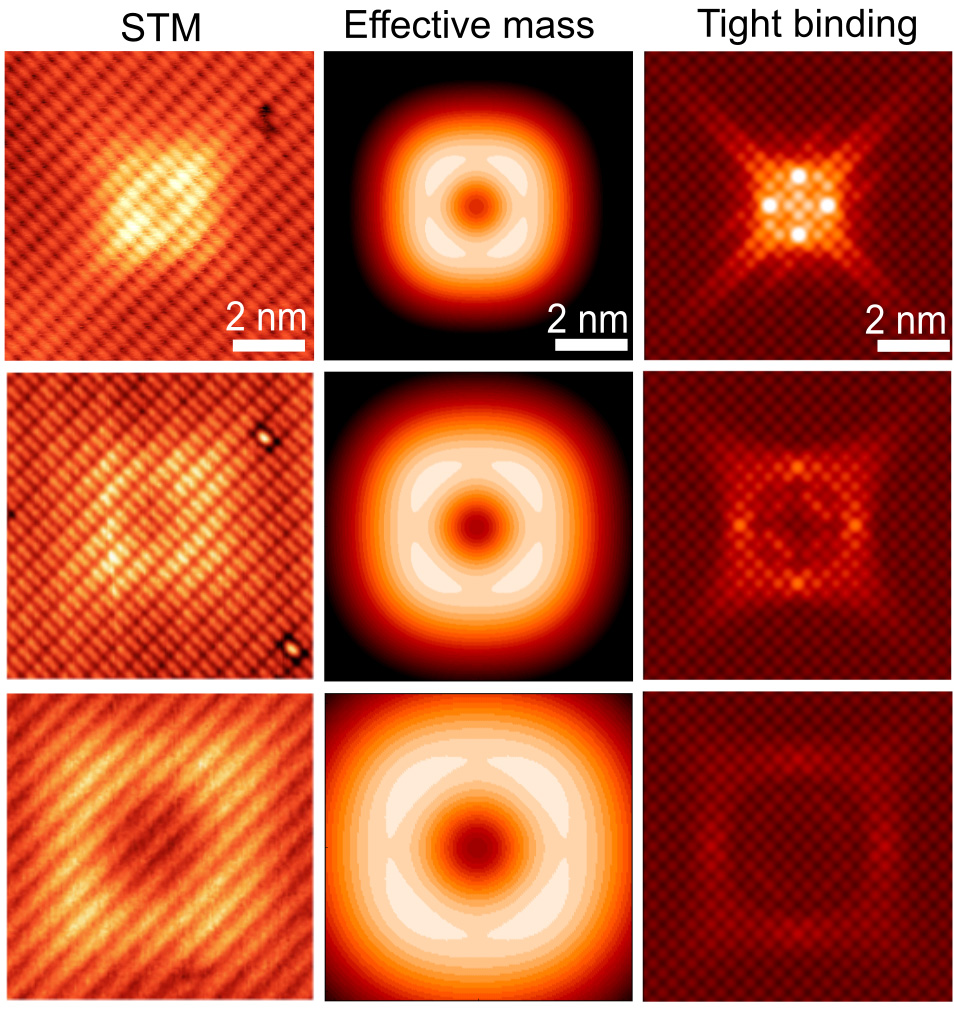}
\caption{Comparison of STM measurements (left column), effective mass modeling (center column), and tight-binding calculations (right column) for acceptor states in silicon. STM image parameters (top to bottom): 1.1~V, 20 pA, z-range 160~pm; 1.6~V, 20~pA, z-range 30~pm; 1.0~V, 5 pA, z-range 100~pm. For the effective mass results, contour plots are shown for acceptors at depths of 2.9, 3.9, and 5.9~nm below the image plane. The tight-binding simulations show topographical scans of the acceptor local density of states at depths of 0.81, 1.62, and 2.17 nm below the surface. Both effective mass and tight binding images use a color intensity scale with a range of 80~pm.}
\label{fig3}
\end{figure}

As the number of atomic planes from the impurity increases, corresponding in the experiment to greater defect depth below the surface, we continue to find good qualitative agreement between STM measurements and EFMA results. Fig.~\ref{fig3} shows a comparison between experimental STM images (left column) and EFMA simulations (center column). The right column of Fig.~\ref{fig3} shows corresponding atomistic tight-binding simulations, which also exhibit excellent agreement with the STM data and are discussed further below. In the STM measurements, the top to bottom rows correspond to increasing feature width that we attribute to defect depth beneath the surface, while in the EFMA simulations they correspond to wave function evaluations at planes located progressively farther from the impurity site (2.9~nm, 3.9~nm, and 5.9~nm). The agreement between STM and EFMA simulations is excellent and becomes increasingly good as the defect depth increases. 

To gain an additional, atomistic perspective on the electronic structure, we turn next to tight‐binding calculations. In particular, the Green’s function method used here is a many‐band single‐particle theory formulated in a TB basis. This method casts the defect problem in terms of bulk Bloch waves scattering off a localized defect potential. Formally, this is expressed using the Dyson equation,
\begin{align}
    \hat{G}(\omega)=(1-\hat{V'}\hat{g}(\omega))^{-1}\hat{g}(\omega)\label{eq:Dyson_equation}
\end{align}
where $V'$ is the perturbative (inhomogeneous) potential associated with the defect, $\hat{g}$ are bulk (unperturbed) Green's functions and $\hat{G}$ are the resultant inhomogeneous Green's functions containing information about the defect resonant energy and wave function. We compute the real-space bulk Green's functions, $g(\omega)$, by taking the inverse Fourier transform of the resolvent of the bulk electronic Hamiltonian, 
\begin{align}
    g(\mathbf{r},\mathbf{r}';z)=\int_{BZ}[z-\hat{H}(\mathbf{k})]^{-1}e^{i(\mathbf{r}-\mathbf{r}')\cdot\mathbf{k}}d\mathbf{k}.
\end{align}

We employ an empirical spds* TB model (including s, p, d, and excited s* orbitals) for silicon that includes spin-orbit coupling~\cite{PhysRevB.57.6493}. This model accurately reproduces the conduction and valence band edges and effective masses throughout the Brillouin zone.

A key detail, as it pertains to the computation cost of this method, is that if $V'$ is well described by a relatively small cluster of atoms with short-ranged interactions, then Eq.~\ref{eq:Dyson_equation} can be efficiently and exactly solved. Expressed in block form, it can be written as,
\begin{align}
        \hat{G}= 
    \begin{pmatrix}
    \hat{G}_{nn}&\hat{G}_{nf}\\
    \hat{G}_{fn}&\hat{G}_{ff}
    \end{pmatrix}=
    \begin{pmatrix}
        \hat{M}_{nn}\hat{g}_{nn}&\hat{M}_{nn}\hat{g}_{nf}\\
        \hat{g}_{fn}\hat{M}_{nn}&\hat{g}_{ff}+\hat{g}_{fn}\hat{V'}_{nn}\hat{M}_{nn}\hat{g}_{nf}
    \end{pmatrix}\label{eq:G_solved}
\end{align}
where $\hat{M}_{n,n}=(1-\hat{g}_{n,n}\hat{V'}_{n,n})^{-1}$. Here the subscripts ``n'' and ``f'' refer to the so-called near field and far field as described in the Supporting Information. In Eq. \ref{eq:G_solved},the computation of the inhomogeneous Green's functions depends only on the local propagator ($g_{ii}$), the propagator from the local site to the defect impurity ($g_{i,n}$), and the description of the local impurity in the near-field defined by $M_{nn}$. As there is no coupling of terms between two differing far-field sites, the defect calculation avoids  truncation effects when computing finite size fields of view.

The local density of states, $\eta$, can be directly extracted from the inhomogeneous Green's functions via the relation,
\begin{align}
    \eta(\mathbf{r};\omega)=\frac{-1}{\pi}\text{Im}[\text{Tr}[\hat{G}(\mathbf{r},\mathbf{r};\omega)]]\,.
\end{align}
Topographical images for comparison with STM images are generated by computing $\eta$ in conjunction with Eq.~\ref{eq:eq_height} for every atomic site in the field of view for a given plane, and then convolving with a spherical Gaussian of full width at half maximum (FWHM) of one-quarter the nearest-neighbor bond distance ($d_{NN}=2.35$\AA).

The right column of Fig.~\ref{fig3} shows acceptor states generated using this method. We found excellent agreement between our TB Green's function calculations for a defect modeled as an ideal substitutional impurity with $T_d$ site symmetry, located 30~meV above the valence band edge, and the experimentally observed acceptor state wave function. The series of images with increasing depth in Fig.~\ref{fig3} demonstrate the evolution of the defect wave function with depth. As with the EFMA descrptions (centre column, Fig.~\ref{fig3}), these images reproduce extremely well the features seen in the experimental data (left column, Fig.~\ref{fig3}), although we obtain slightly different predictions for the defect depth (0.81, 1.62, and 2.17 nm) compared to the EFMA theory.

We note that the EFMA and TB models have different strengths, which can account for the discrepancies between them. The simpler (isotropic bulk dispersion) effective mass model permits a straightforward modification of the valence effective masses to study the effects on the wave function shape and choose optimal parameters. The tight-binding model includes spherical anisotropy of the dispersion, however features like the effective mass depend on many parameters that also affect other properties, so adjusting one mass is very difficult. Neither model includes the distortion of the electronic structure induced by the surface, which may also have an effect on the correct masses and dispersion anisotropy.

To assess the possibility that the observed acceptor state could originate from a microscopic defect with a symmetry different from that of the silicon crystal lattice, we modeled a split-vacancy configuration characterized by $S_6$ site symmetry (see Supporting Information Section~2.4). The simulated wave function for this defect exhibited strong deviations from the experimentally observed square ring-like structure. We therefore conclude that the split-vacancy model is inconsistent with the experimental data and is an unlikely candidate for the observed defect.

This work presents the first images of acceptor state wave functions in silicon. Through scanning tunneling microscopy measurements, the acceptor states are identified as square ring-like features with strong electronic contrast superimposed on the atomic lattice of the Si(001) surface. Tunneling spectroscopy and spatial \didv\ conductance maps confirm their spatial structure, energetic position approximately 30~meV above the valence band edge, and their role in inducing a $p$-type Si(001) surface. Comparison with effective mass and atomistic tight-binding calculations confirms that the observed features are consistent with acceptor states in silicon and provides deeper insight into their characteristics. The effective mass model clearly reveals a $d$-like contribution to the square ring appearance, while the tight-binding calculations indicate that the observed symmetry arises from a substitutional impurity with $T_d$ site symmetry. The combination of high-resolution experiment and detailed theoretical modeling demonstrated here provides the foundation for a thorough understanding of acceptor states in silicon, essential for future quantum technological applications based on these defects.

The authors acknowledge useful conversations with A. J. Fisher, S. K. Clowes, and J. A. Gupta. The STM experiments were financially supported by Engineering \& Physical Sciences Research Council (EP/L002140/1, EP/M009564/1); effective mass simulations by a Marie Sklodowska-Curie Grant 956548; tight-binding simulations by the U. S. Air Force Office of Scientific Research under Award number FA9550-24-1-0355; code development enabling tight-binding simulations was supported by the U.S. Department of Energy, Office of Basic Energy Sciences, under Award No. DE-SC0016379.


\begin{mcitethebibliography}{38}
\providecommand*\natexlab[1]{#1}
\providecommand*\mciteSetBstSublistMode[1]{}
\providecommand*\mciteSetBstMaxWidthForm[2]{}
\providecommand*\mciteBstWouldAddEndPuncttrue
  {\def\EndOfBibitem{\unskip.}}
\providecommand*\mciteBstWouldAddEndPunctfalse
  {\let\EndOfBibitem\relax}
\providecommand*\mciteSetBstMidEndSepPunct[3]{}
\providecommand*\mciteSetBstSublistLabelBeginEnd[3]{}
\providecommand*\EndOfBibitem{}
\mciteSetBstSublistMode{f}
\mciteSetBstMaxWidthForm{subitem}{(\alph{mcitesubitemcount})}
\mciteSetBstSublistLabelBeginEnd
  {\mcitemaxwidthsubitemform\space}
  {\relax}
  {\relax}

\bibitem[{Yu, Pater} and Cardona(2010){Yu, Pater}, and Cardona]{YuCardona}
{Yu, Pater},~Y.; Cardona,~M. \emph{{Fundamentals of Semiconductors}}, 4th ed.;
  Springer, 2010\relax
\mciteBstWouldAddEndPuncttrue
\mciteSetBstMidEndSepPunct{\mcitedefaultmidpunct}
{\mcitedefaultendpunct}{\mcitedefaultseppunct}\relax
\EndOfBibitem
\bibitem[Muhonen \latin{et~al.}(2014)Muhonen, Dehollain, Laucht, Hudson, Kalra,
  Sekiguchi, Itoh, Jamieson, McCallum, Dzurak, and Morello]{Muhonen2014}
Muhonen,~J.~T.; Dehollain,~J.~P.; Laucht,~A.; Hudson,~F.~E.; Kalra,~R.;
  Sekiguchi,~T.; Itoh,~K.~M.; Jamieson,~D.~N.; McCallum,~J.~C.; Dzurak,~A.~S.;
  Morello,~A. {Storing quantum information for 30 seconds in a nanoelectronic
  device}. \emph{Nature Nanotechnology} \textbf{2014}, \emph{9}, 986--991\relax
\mciteBstWouldAddEndPuncttrue
\mciteSetBstMidEndSepPunct{\mcitedefaultmidpunct}
{\mcitedefaultendpunct}{\mcitedefaultseppunct}\relax
\EndOfBibitem
\bibitem[Kobayashi \latin{et~al.}(2020)Kobayashi, Salfi, Chua, van~der Heijden,
  House, Culcer, Hutchison, Johnson, McCallum, Riemann, Abrosimov, Becker,
  Pohl, Simmons, and Rogge]{Kobayashi2020}
Kobayashi,~T.; Salfi,~J.; Chua,~C.; van~der Heijden,~J.; House,~M.~G.;
  Culcer,~D.; Hutchison,~W.~D.; Johnson,~B.~C.; McCallum,~J.~C.; Riemann,~H.;
  Abrosimov,~N.~V.; Becker,~P.; Pohl,~H.~J.; Simmons,~M.~Y.; Rogge,~S.
  Engineering long spin coherence times of spin–orbit qubits in silicon.
  \emph{Nature Materials} \textbf{2020}, \emph{20}, 38--42\relax
\mciteBstWouldAddEndPuncttrue
\mciteSetBstMidEndSepPunct{\mcitedefaultmidpunct}
{\mcitedefaultendpunct}{\mcitedefaultseppunct}\relax
\EndOfBibitem
\bibitem[Weber \latin{et~al.}(2010)Weber, Koehl, Varley, Janotti, Buckley, {Van
  de Walle}, and Awschalom]{Weber2010}
Weber,~J.~R.; Koehl,~W.~F.; Varley,~J.~B.; Janotti,~A.; Buckley,~B.~B.; {Van de
  Walle},~C.~G.; Awschalom,~D.~D. {Quantum computing with defects.} \emph{Proc.
  Natl. Acad. Sci. USA} \textbf{2010}, \emph{107}, 8513--8518\relax
\mciteBstWouldAddEndPuncttrue
\mciteSetBstMidEndSepPunct{\mcitedefaultmidpunct}
{\mcitedefaultendpunct}{\mcitedefaultseppunct}\relax
\EndOfBibitem
\bibitem[Koenraad and Flatt{\'{e}}(2011)Koenraad, and
  Flatt{\'{e}}]{Koenraad2011}
Koenraad,~P.~M.; Flatt{\'{e}},~M.~E. {Single dopants in semiconductors.}
  \emph{Nat. Mater.} \textbf{2011}, \emph{10}, 91--100\relax
\mciteBstWouldAddEndPuncttrue
\mciteSetBstMidEndSepPunct{\mcitedefaultmidpunct}
{\mcitedefaultendpunct}{\mcitedefaultseppunct}\relax
\EndOfBibitem
\bibitem[Schofield \latin{et~al.}(2003)Schofield, Curson, Simmons, Ruess,
  Hallam, Oberbeck, and Clark]{Schofield2003}
Schofield,~S.~R.; Curson,~N.~J.; Simmons,~M.~Y.; Ruess,~F.~J.; Hallam,~T.;
  Oberbeck,~L.; Clark,~R.~G. {Atomically Precise Placement of Single Dopants in
  Si}. \emph{Phys. Rev. Lett.} \textbf{2003}, \emph{91}, 136104\relax
\mciteBstWouldAddEndPuncttrue
\mciteSetBstMidEndSepPunct{\mcitedefaultmidpunct}
{\mcitedefaultendpunct}{\mcitedefaultseppunct}\relax
\EndOfBibitem
\bibitem[Stock \latin{et~al.}(2020)Stock, Warschkow, Constantinou, Li, Fearn,
  Crane, Hofmann, K{\"{o}}lker, Mckenzie, Schofield, and Curson]{Stock2020}
Stock,~T.~J.; Warschkow,~O.; Constantinou,~P.~C.; Li,~J.; Fearn,~S.; Crane,~E.;
  Hofmann,~E.~V.; K{\"{o}}lker,~A.; Mckenzie,~D.~R.; Schofield,~S.~R.;
  Curson,~N.~J. {Atomic-Scale Patterning of Arsenic in Silicon by Scanning
  Tunneling Microscopy}. \emph{ACS Nano} \textbf{2020}, \emph{14},
  3316--3327\relax
\mciteBstWouldAddEndPuncttrue
\mciteSetBstMidEndSepPunct{\mcitedefaultmidpunct}
{\mcitedefaultendpunct}{\mcitedefaultseppunct}\relax
\EndOfBibitem
\bibitem[Stock \latin{et~al.}(2024)Stock, Warschkow, Constantinou, Bowler,
  Schofield, and Curson]{Stock2024}
Stock,~T. J.~Z.; Warschkow,~O.; Constantinou,~P.~C.; Bowler,~D.~R.;
  Schofield,~S.~R.; Curson,~N.~J. {Single‐Atom Control of Arsenic
  Incorporation in Silicon for High‐Yield Artificial Lattice Fabrication}.
  \emph{Advanced Materials} \textbf{2024}, \emph{23}, 12282\relax
\mciteBstWouldAddEndPuncttrue
\mciteSetBstMidEndSepPunct{\mcitedefaultmidpunct}
{\mcitedefaultendpunct}{\mcitedefaultseppunct}\relax
\EndOfBibitem
\bibitem[{\v{S}}kereň \latin{et~al.}(2020){\v{S}}kereň, K{\"{o}}ster,
  Douhard, Fleischmann, and Fuhrer]{Skeren2020}
{\v{S}}kereň,~T.; K{\"{o}}ster,~S.~A.; Douhard,~B.; Fleischmann,~C.;
  Fuhrer,~A. {Bipolar device fabrication using a scanning tunnelling
  microscope}. \emph{Nat. Electron.} \textbf{2020}, \emph{3}, 524--530\relax
\mciteBstWouldAddEndPuncttrue
\mciteSetBstMidEndSepPunct{\mcitedefaultmidpunct}
{\mcitedefaultendpunct}{\mcitedefaultseppunct}\relax
\EndOfBibitem
\bibitem[Fuechsle \latin{et~al.}(2012)Fuechsle, Miwa, Mahapatra, Ryu, Lee,
  Warschkow, Hollenberg, Klimeck, and Simmons]{Fuechsle2012}
Fuechsle,~M.; Miwa,~J.~A.; Mahapatra,~S.; Ryu,~H.; Lee,~S.; Warschkow,~O.;
  Hollenberg,~L. C.~L.; Klimeck,~G.; Simmons,~M.~Y. {A single-atom transistor.}
  \emph{Nat. Nano.} \textbf{2012}, \emph{7}, 242\relax
\mciteBstWouldAddEndPuncttrue
\mciteSetBstMidEndSepPunct{\mcitedefaultmidpunct}
{\mcitedefaultendpunct}{\mcitedefaultseppunct}\relax
\EndOfBibitem
\bibitem[He \latin{et~al.}(2019)He, Gorman, Keith, Kranz, Keizer, and
  Simmons]{He2019}
He,~Y.; Gorman,~S.~K.; Keith,~D.; Kranz,~L.; Keizer,~J.~G.; Simmons,~M.~Y. {A
  two-qubit gate between phosphorus donor electrons in silicon}. \emph{Nature}
  \textbf{2019}, \emph{571}, 371--375\relax
\mciteBstWouldAddEndPuncttrue
\mciteSetBstMidEndSepPunct{\mcitedefaultmidpunct}
{\mcitedefaultendpunct}{\mcitedefaultseppunct}\relax
\EndOfBibitem
\bibitem[Thorvaldson \latin{et~al.}(2025)Thorvaldson, Poulos, Moehle, Misha,
  Edlbauer, Reiner, Geng, Voisin, Jones, Donnelly, \latin{et~al.}
  others]{Thorvaldson2024}
Thorvaldson,~I.; Poulos,~D.; Moehle,~C.; Misha,~S.; Edlbauer,~H.; Reiner,~J.;
  Geng,~H.; Voisin,~B.; Jones,~M.; Donnelly,~M.; others Grover’s algorithm in
  a four-qubit silicon processor above the fault-tolerant threshold.
  \emph{Nature Nanotechnology} \textbf{2025}, \emph{20}, 472–477\relax
\mciteBstWouldAddEndPuncttrue
\mciteSetBstMidEndSepPunct{\mcitedefaultmidpunct}
{\mcitedefaultendpunct}{\mcitedefaultseppunct}\relax
\EndOfBibitem
\bibitem[Hill \latin{et~al.}(2015)Hill, Peretz, Hile, House, Fuechsle, Rogge,
  Simmons, and Hollenberg]{Hill2015a}
Hill,~C.~D.; Peretz,~E.; Hile,~S.~J.; House,~M.~G.; Fuechsle,~M.; Rogge,~S.;
  Simmons,~M.~Y.; Hollenberg,~L.~C. {Quantum Computing: A surface code quantum
  computer in silicon}. \emph{Sci. Adv.} \textbf{2015}, \emph{1},
  e1500707\relax
\mciteBstWouldAddEndPuncttrue
\mciteSetBstMidEndSepPunct{\mcitedefaultmidpunct}
{\mcitedefaultendpunct}{\mcitedefaultseppunct}\relax
\EndOfBibitem
\bibitem[Ebert(2003)]{Ebert2003}
Ebert,~H. Imaging defects and dopants. \emph{Materials Today} \textbf{2003},
  \emph{6}, 36--43\relax
\mciteBstWouldAddEndPuncttrue
\mciteSetBstMidEndSepPunct{\mcitedefaultmidpunct}
{\mcitedefaultendpunct}{\mcitedefaultseppunct}\relax
\EndOfBibitem
\bibitem[Tang and Flatt\'e(2004)Tang, and Flatt\'e]{PhysRevLett.92.047201}
Tang,~J.-M.; Flatt\'e,~M.~E. Multiband Tight-Binding Model of Local Magnetism
  in
  ${\mathrm{G}\mathrm{a}}_{1\ensuremath{-}x}{\mathrm{M}\mathrm{n}}_{x}\mathrm{A}\mathrm{s}$.
  \emph{Phys. Rev. Lett.} \textbf{2004}, \emph{92}, 047201\relax
\mciteBstWouldAddEndPuncttrue
\mciteSetBstMidEndSepPunct{\mcitedefaultmidpunct}
{\mcitedefaultendpunct}{\mcitedefaultseppunct}\relax
\EndOfBibitem
\bibitem[Yakunin \latin{et~al.}(2004)Yakunin, Silov, Koenraad, Wolter, {Van
  Roy}, {De Boeck}, Tang, and Flatt{\'{e}}]{Yakunin2004a}
Yakunin,~A.; Silov,~A.~Y.; Koenraad,~P.~M.; Wolter,~J.; {Van Roy},~W.; {De
  Boeck},~J.; Tang,~J.-M.; Flatt{\'{e}},~M.~E. {Spatial Structure of an
  Individual Mn Acceptor in GaAs}. \emph{Phys. Rev. Lett.} \textbf{2004},
  \emph{92}, 216806\relax
\mciteBstWouldAddEndPuncttrue
\mciteSetBstMidEndSepPunct{\mcitedefaultmidpunct}
{\mcitedefaultendpunct}{\mcitedefaultseppunct}\relax
\EndOfBibitem
\bibitem[Marczinowski \latin{et~al.}(2007)Marczinowski, Wiebe, Tang,
  Flatt{\'{e}}, Meier, Morgenstern, and Wiesendanger]{Marczinowski2007}
Marczinowski,~F.; Wiebe,~J.; Tang,~J.~M.; Flatt{\'{e}},~M.~E.; Meier,~F.;
  Morgenstern,~M.; Wiesendanger,~R. {Local electronic structure near Mn
  acceptors in InAs: Surface-induced symmetry breaking and coupling to host
  states}. \emph{Phys. Rev. Lett.} \textbf{2007}, \emph{99}, 10--13\relax
\mciteBstWouldAddEndPuncttrue
\mciteSetBstMidEndSepPunct{\mcitedefaultmidpunct}
{\mcitedefaultendpunct}{\mcitedefaultseppunct}\relax
\EndOfBibitem
\bibitem[Loth \latin{et~al.}(2006)Loth, Wenderoth, Winking, Ulbrich, Malzer,
  and D{\"{o}}hler]{Loth2006}
Loth,~S.; Wenderoth,~M.; Winking,~L.; Ulbrich,~R.; Malzer,~S.; D{\"{o}}hler,~G.
  {Probing Semiconductor Gap States with Resonant Tunneling}. \emph{Phys. Rev.
  Lett.} \textbf{2006}, \emph{96}, 066403\relax
\mciteBstWouldAddEndPuncttrue
\mciteSetBstMidEndSepPunct{\mcitedefaultmidpunct}
{\mcitedefaultendpunct}{\mcitedefaultseppunct}\relax
\EndOfBibitem
\bibitem[Garleff \latin{et~al.}(2010)Garleff, Wijnheijmer, Silov, van Bree,
  {Van Roy}, Tang, Flatt{\'{e}}, and Koenraad]{Garleff2010}
Garleff,~J.~K.; Wijnheijmer,~A.; Silov,~A.; van Bree,~J.; {Van Roy},~W.;
  Tang,~J.-M.; Flatt{\'{e}},~M.~E.; Koenraad,~P.~M. {Enhanced binding energy of
  manganese acceptors close to the GaAs(110) surface}. \emph{Phys. Rev. B}
  \textbf{2010}, \emph{82}, 035303\relax
\mciteBstWouldAddEndPuncttrue
\mciteSetBstMidEndSepPunct{\mcitedefaultmidpunct}
{\mcitedefaultendpunct}{\mcitedefaultseppunct}\relax
\EndOfBibitem
\bibitem[Teichmann \latin{et~al.}(2008)Teichmann, Wenderoth, Loth, Ulbrich,
  Garleff, Wijnheijmer, and Koenraad]{Teichmann2008}
Teichmann,~K.; Wenderoth,~M.; Loth,~S.; Ulbrich,~R.; Garleff,~J.;
  Wijnheijmer,~A.; Koenraad,~P.~M. {Controlled charge switching on a single
  donor with a scanning tunneling microscope}. \emph{Phys. Rev. Lett.}
  \textbf{2008}, \emph{101}, 076103\relax
\mciteBstWouldAddEndPuncttrue
\mciteSetBstMidEndSepPunct{\mcitedefaultmidpunct}
{\mcitedefaultendpunct}{\mcitedefaultseppunct}\relax
\EndOfBibitem
\bibitem[Studer \latin{et~al.}(2012)Studer, Br{\'{a}}zdov{\'{a}}, Schofield,
  Bowler, Hirjibehedin, and Curson]{Studer2012}
Studer,~P.; Br{\'{a}}zdov{\'{a}},~V.; Schofield,~S.~R.; Bowler,~D.~R.;
  Hirjibehedin,~C.~F.; Curson,~N.~J. {Site-dependent ambipolar charge states
  induced by group {V} atoms in a silicon surface.} \emph{ACS Nano}
  \textbf{2012}, \emph{6}, 10456--10462\relax
\mciteBstWouldAddEndPuncttrue
\mciteSetBstMidEndSepPunct{\mcitedefaultmidpunct}
{\mcitedefaultendpunct}{\mcitedefaultseppunct}\relax
\EndOfBibitem
\bibitem[Liu \latin{et~al.}(2001)Liu, Yu, and Lyding]{Liu2001}
Liu,~L.; Yu,~J.; Lyding,~J.~W. {Atom-resolved three-dimensional mapping of
  boron dopants in Si(100) by scanning tunneling microscopy}. \emph{Appl. Phys.
  Lett.} \textbf{2001}, \emph{78}, 386--388\relax
\mciteBstWouldAddEndPuncttrue
\mciteSetBstMidEndSepPunct{\mcitedefaultmidpunct}
{\mcitedefaultendpunct}{\mcitedefaultseppunct}\relax
\EndOfBibitem
\bibitem[Liu \latin{et~al.}(2002)Liu, Yu, and Lyding]{liu2002a}
Liu,~L.; Yu,~J.; Lyding,~J.~W. {Subsurface Dopant-Induced Features on the
  Si(100)2$\times$1:H Surface: Fundamental the Study and Applications}.
  \emph{{IEEE} Trans. Nanotechnol.} \textbf{2002}, \emph{1}, 176--183\relax
\mciteBstWouldAddEndPuncttrue
\mciteSetBstMidEndSepPunct{\mcitedefaultmidpunct}
{\mcitedefaultendpunct}{\mcitedefaultseppunct}\relax
\EndOfBibitem
\bibitem[Nishizawa \latin{et~al.}(2008)Nishizawa, Bolotov, and
  Kanayama]{Nishizawa2008}
Nishizawa,~M.; Bolotov,~L.; Kanayama,~T. {Scanning tunneling
  microscopy/spectroscopy as an atomic-resolution dopant profiling method in
  Si}. \emph{J. Phys. Conf. Ser.} \textbf{2008}, \emph{106}, 012002\relax
\mciteBstWouldAddEndPuncttrue
\mciteSetBstMidEndSepPunct{\mcitedefaultmidpunct}
{\mcitedefaultendpunct}{\mcitedefaultseppunct}\relax
\EndOfBibitem
\bibitem[Miwa \latin{et~al.}(2013)Miwa, Hofmann, Simmons, and Wells]{Miwa2013}
Miwa,~J.~A.; Hofmann,~P.; Simmons,~M.~Y.; Wells,~J.~W. {Direct measurement of
  the band structure of a buried two-dimensional electron gas}. \emph{Phys.
  Rev. Lett.} \textbf{2013}, \emph{110}, 136801\relax
\mciteBstWouldAddEndPuncttrue
\mciteSetBstMidEndSepPunct{\mcitedefaultmidpunct}
{\mcitedefaultendpunct}{\mcitedefaultseppunct}\relax
\EndOfBibitem
\bibitem[Mol \latin{et~al.}(2013)Mol, Salfi, Miwa, Simmons, and Rogge]{Mol2013}
Mol,~J.~A.; Salfi,~J.; Miwa,~J.~A.; Simmons,~M.~Y.; Rogge,~S. {Interplay
  between quantum confinement and dielectric mismatch for ultrashallow
  dopants}. \emph{Phys. Rev. B} \textbf{2013}, \emph{87}, 245417\relax
\mciteBstWouldAddEndPuncttrue
\mciteSetBstMidEndSepPunct{\mcitedefaultmidpunct}
{\mcitedefaultendpunct}{\mcitedefaultseppunct}\relax
\EndOfBibitem
\bibitem[Sinthiptharakoon \latin{et~al.}(2014)Sinthiptharakoon, Schofield,
  Studer, Br{\'{a}}zdov{\'{a}}, Hirjibehedin, Bowler, and
  Curson]{Sinthiptharakoon2014}
Sinthiptharakoon,~K.; Schofield,~S.~R.; Studer,~P.; Br{\'{a}}zdov{\'{a}},~V.;
  Hirjibehedin,~C.~F.; Bowler,~D.~R.; Curson,~N.~J. {Investigating individual
  arsenic dopant atoms in silicon using low-temperature scanning tunnelling
  microscopy}. \emph{Journal of Physics Condensed Matter} \textbf{2014},
  \emph{26}, 012001\relax
\mciteBstWouldAddEndPuncttrue
\mciteSetBstMidEndSepPunct{\mcitedefaultmidpunct}
{\mcitedefaultendpunct}{\mcitedefaultseppunct}\relax
\EndOfBibitem
\bibitem[Usman \latin{et~al.}(2016)Usman, Bocquel, Salfi, Voisin, Tankasala,
  Rahman, Simmons, Rogge, and Hollenberg]{Usman2016}
Usman,~M.; Bocquel,~J.; Salfi,~J.; Voisin,~B.; Tankasala,~A.; Rahman,~R.;
  Simmons,~M.~Y.; Rogge,~S.; Hollenberg,~L.~C. {Spatial metrology of dopants in
  silicon with exact lattice site precision}. \emph{Nat. Nanotechnol.}
  \textbf{2016}, \emph{11}, 763--768\relax
\mciteBstWouldAddEndPuncttrue
\mciteSetBstMidEndSepPunct{\mcitedefaultmidpunct}
{\mcitedefaultendpunct}{\mcitedefaultseppunct}\relax
\EndOfBibitem
\bibitem[Brazdova \latin{et~al.}(2017)Brazdova, Bowler, Sinthiptharakoon,
  Studer, Rahnejat, Curson, Schofield, and Fisher]{Brazdova2015}
Brazdova,~V.; Bowler,~D.~R.; Sinthiptharakoon,~K.; Studer,~P.; Rahnejat,~A.;
  Curson,~N.~J.; Schofield,~S.~R.; Fisher,~A.~J. {Exact location of dopants
  below the Si(001):H surface from scanning tunneling microscopy and density
  functional theory}. \emph{Physical Review B} \textbf{2017}, \emph{95},
  075408\relax
\mciteBstWouldAddEndPuncttrue
\mciteSetBstMidEndSepPunct{\mcitedefaultmidpunct}
{\mcitedefaultendpunct}{\mcitedefaultseppunct}\relax
\EndOfBibitem
\bibitem[Salfi \latin{et~al.}(2014)Salfi, Mol, Rahman, Klimeck, Simmons,
  Hollenberg, and Rogge]{Salfi2014}
Salfi,~J.; Mol,~J.~A.; Rahman,~R.; Klimeck,~G.; Simmons,~M.~Y.; Hollenberg,~L.
  C.~L.; Rogge,~S. {Spatially resolving valley quantum interference of a donor
  in silicon}. \emph{Nat. Mater.} \textbf{2014}, \emph{13}, 605--610\relax
\mciteBstWouldAddEndPuncttrue
\mciteSetBstMidEndSepPunct{\mcitedefaultmidpunct}
{\mcitedefaultendpunct}{\mcitedefaultseppunct}\relax
\EndOfBibitem
\bibitem[Voisin \latin{et~al.}(2020)Voisin, Bocquel, Tankasala, Usman, Salfi,
  Rahman, Simmons, Hollenberg, and Rogge]{Voisin2020}
Voisin,~B.; Bocquel,~J.; Tankasala,~A.; Usman,~M.; Salfi,~J.; Rahman,~R.;
  Simmons,~M.~Y.; Hollenberg,~L.~C.; Rogge,~S. {Valley interference and spin
  exchange at the atomic scale in silicon}. \emph{Nat. Commun.} \textbf{2020},
  \emph{11}, 1--11\relax
\mciteBstWouldAddEndPuncttrue
\mciteSetBstMidEndSepPunct{\mcitedefaultmidpunct}
{\mcitedefaultendpunct}{\mcitedefaultseppunct}\relax
\EndOfBibitem
\bibitem[Peach \latin{et~al.}(2018)Peach, Homewood, Lourenco, Hughes, Saeedi,
  Stavrias, Li, Chick, Murdin, and Clowes]{Peach2018}
Peach,~T.; Homewood,~K.; Lourenco,~M.; Hughes,~M.; Saeedi,~K.; Stavrias,~N.;
  Li,~J.; Chick,~S.; Murdin,~B.; Clowes,~S. {The Effect of Lattice Damage and
  Annealing Conditions on the Hyperfine Structure of Ion Implanted Bismuth
  Donors in Silicon}. \emph{Adv. Quantum Technol.} \textbf{2018}, \emph{1},
  1--6\relax
\mciteBstWouldAddEndPuncttrue
\mciteSetBstMidEndSepPunct{\mcitedefaultmidpunct}
{\mcitedefaultendpunct}{\mcitedefaultseppunct}\relax
\EndOfBibitem
\bibitem[Yakunin \latin{et~al.}(2007)Yakunin, Silov, Koenraad, Tang,
  Flatt{\'{e}}, Primus, {Van Roy}, {De Boeck}, Monakhov, Romanov, Panaiotti,
  and Averkiev]{Yakunin2007}
Yakunin,~A.~M.; Silov,~A.~Y.; Koenraad,~P.~M.; Tang,~J.~M.;
  Flatt{\'{e}},~M.~E.; Primus,~J.~L.; {Van Roy},~W.; {De Boeck},~J.;
  Monakhov,~A.~M.; Romanov,~K.~S.; Panaiotti,~I.~E.; Averkiev,~N.~S. {Warping a
  single Mn acceptor wavefunction by straining the GaAs host}. \emph{Nature
  Materials} \textbf{2007}, \emph{6}, 512--515\relax
\mciteBstWouldAddEndPuncttrue
\mciteSetBstMidEndSepPunct{\mcitedefaultmidpunct}
{\mcitedefaultendpunct}{\mcitedefaultseppunct}\relax
\EndOfBibitem
\bibitem[Averkiev and Il'Inskii(1994)Averkiev, and Il'Inskii]{averkiev1994spin}
Averkiev,~N.; Il'Inskii,~S.~Y. Spin ordering of carriers localized at two deep
  centers in cubic semiconductors. \emph{Physics of the Solid State}
  \textbf{1994}, \emph{36}, 278--283\relax
\mciteBstWouldAddEndPuncttrue
\mciteSetBstMidEndSepPunct{\mcitedefaultmidpunct}
{\mcitedefaultendpunct}{\mcitedefaultseppunct}\relax
\EndOfBibitem
\bibitem[Vinh \latin{et~al.}(2013)Vinh, Redlich, van~der Meer, Pidgeon,
  Greenland, Lynch, Aeppli, and Murdin]{PhysRevX.3.011019}
Vinh,~N.~Q.; Redlich,~B.; van~der Meer,~A. F.~G.; Pidgeon,~C.~R.;
  Greenland,~P.~T.; Lynch,~S.~A.; Aeppli,~G.; Murdin,~B.~N. Time-Resolved
  Dynamics of Shallow Acceptor Transitions in Silicon. \emph{Phys. Rev. X}
  \textbf{2013}, \emph{3}, 011019\relax
\mciteBstWouldAddEndPuncttrue
\mciteSetBstMidEndSepPunct{\mcitedefaultmidpunct}
{\mcitedefaultendpunct}{\mcitedefaultseppunct}\relax
\EndOfBibitem
\bibitem[Baldereschi and Lipari(1974)Baldereschi, and Lipari]{Baldereschi1974}
Baldereschi,~A.; Lipari,~N.~O. {Cubic contributions to the spherical model of
  shallow acceptor states}. \emph{Phys. Rev. B} \textbf{1974}, \emph{9},
  1525--1539\relax
\mciteBstWouldAddEndPuncttrue
\mciteSetBstMidEndSepPunct{\mcitedefaultmidpunct}
{\mcitedefaultendpunct}{\mcitedefaultseppunct}\relax
\EndOfBibitem
\bibitem[Jancu \latin{et~al.}(1998)Jancu, Scholz, Beltram, and
  Bassani]{PhysRevB.57.6493}
Jancu,~J.-M.; Scholz,~R.; Beltram,~F.; Bassani,~F. Empirical
  ${\mathrm{spds}}^{*}$ tight-binding calculation for cubic semiconductors:
  General method and material parameters. \emph{Phys. Rev. B} \textbf{1998},
  \emph{57}, 6493--6507\relax
\mciteBstWouldAddEndPuncttrue
\mciteSetBstMidEndSepPunct{\mcitedefaultmidpunct}
{\mcitedefaultendpunct}{\mcitedefaultseppunct}\relax
\EndOfBibitem
\end{mcitethebibliography}

\providecommand{\latin}[1]{#1}
\makeatletter
\providecommand{\doi}
  {\begingroup\let\do\@makeother\dospecials
  \catcode`\{=1 \catcode`\}=2 \doi@aux}
\providecommand{\doi@aux}[1]{\endgroup\texttt{#1}}
\makeatother
\providecommand*\mcitethebibliography{\thebibliography}
\csname @ifundefined\endcsname{endmcitethebibliography}
  {\let\endmcitethebibliography\endthebibliography}{}

\clearpage


\section*{Supplementary material}

# Supporting Information for

## *Imaging the acceptor wave function anisotropy in silicon*

Manuel Siegl[1,2], Julian Zanon[3], Joseph Sink[4], Adonai Rodrigues da Cruz[4], Holly Hedgeland[1], Neil J. Curson[1,5], Michael Flatté[*3,4], and Steven R. Schofield[†1,2]

[1]London Centre for Nanotechnology, University College London, WC1H 0AH, London, UK
[2]Department of Physics and Astronomy, University College London, WC1E 6BT, London, UK
[3]Department of Applied Physics and Science Education, Eindhoven University of Technology, Eindhoven 5612 AZ, The Netherlands
[4]Department of Physics and Astronomy, University of Iowa, Iowa City, Iowa 52242, USA
[5]Department of Electronic and Electrical Engineering, University College London, WC1E 6BT, London, UK

May 14, 2025

This supplement material is organized as follows: Additional experimental methods and data are presented in Sec. S1., the effective mass approximation is discussed in Sec. S2.1., and the tight-binding method follows in Sec. S2.3.. The data presented in Sec. S1. is extensively discussed in Siegl's PhD thesis [5].

## S1. Supplementary experimental methods and results

Ion implantation was performed at the University of Surrey using $Bi^+$ ions at five energies between 0.25 and 2.0 MeV, with doses ranging from 0.41 to $1.9 \times 10^{15}$ cm$^{-2}$. The implantation parameters are listed in Table S1.

The depth profile resulting from this implantation sequence was estimated using the Stopping and Range of Ions in Matter (SRIM) simulation package [7], as shown in Supplementary Fig. S1. The simulation predicts a rapid increase in Bi concentration from the surface to approximately $1 \times 10^{20}$ cm$^{-3}$ at 20 nm depth, remaining nearly constant to a depth of 550 nm, and then decaying toward background by 700 nm.

---

[*]Email: `michaelflatte@quantumsci.net`
[†]Email: `s.schofield@ucl.ac.uk`

Table S1: Ion implantation parameters for bismuth in Si(001).

| Energy (MeV) | Dose (cm$^{-2}$) |
|---:|---:|
| 2.00 | $1.9 \times 10^{15}$ |
| 1.35 | $1.2 \times 10^{15}$ |
| 0.85 | $0.9 \times 10^{15}$ |
| 0.50 | $0.64 \times 10^{15}$ |
| 0.25 | $0.41 \times 10^{15}$ |

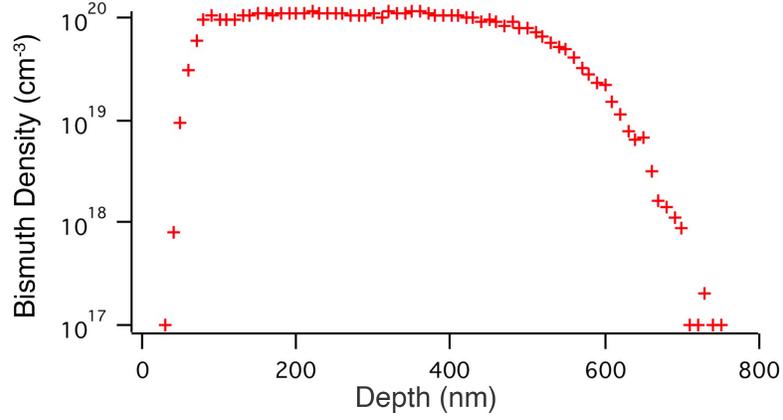

Figure S1: SRIM-estimated depth profile of Bi ion concentration for the implantation conditions listed in Table S1.

To further characterize the behavior of the acceptor states, we performed bias-dependent imaging, as shown in Supplementary Fig. S2a, which shows empty- and filled-state STM images of a single acceptor over a range of imaging biases from ±1.7 V to ±1.0 V. For each bias polarity, the overall appearance of the state remains unchanged within this bias range, with only a decrease in relative intensity compared to the background at higher biases. This behavior is consistent with an increasing contribution of bulk band states to the tunneling current and rules out interpretations based on quasiparticle interference or Friedel oscillations.

We also performed a statistical analysis of the spatial characteristics of many features. To quantify the spatial characteristics of the acceptor states, we extracted line profiles along the [110] direction, as illustrated in Supplementary Fig. S2b. The width and intensity of 75 individual acceptor states were measured and are summarized in Supplementary Fig. S2c. The data reveal an exponential decay of intensity with increasing feature width, consistent with a distribution of acceptor depths: deeper states exhibit reduced spatial overlap with the STM tip wave function and a broader lateral extent at the surface. These results support our interpretation that the variability in the observed features arises primarily from differences in the depth of the acceptor states beneath the surface.

## S2. Supplementary theoretical methods and results

### S2.1. Effective mass approximation (EFMA)

Considering the spherical symmetric Luttinger Hamiltonian with the impurity described by a zero-range potential, it is possible to find an analytical solution for an acceptor ground state as shown in [1]. Following [6], the spherical ground state can be used to construct the following solution

$$\Psi_{3/2}(r,\theta,\phi) = c^{\Gamma_1} R_0(r) \begin{pmatrix} Y_{0,0} \\ 0 \\ 0 \\ 0 \end{pmatrix} + \frac{c^{\Gamma_{25}}}{\sqrt{2}} R_2(r) \begin{pmatrix} 0 \\ -2Y_{2,1} \\ Y_{2,2} - Y_{2,-2} \\ 0 \end{pmatrix} + \frac{c^{\Gamma_{12}}}{\sqrt{2}} R_2(r) \begin{pmatrix} \sqrt{2} Y_{2,0} \\ 0 \\ Y_{2,2} + Y_{2,-2} \\ 0 \end{pmatrix}, \tag{1}$$

$$\Psi_{1/2}(r,\theta,\phi) = c^{\Gamma_1} R_0(r) \begin{pmatrix} 0 \\ Y_{0,0} \\ 0 \\ 0 \end{pmatrix} + \frac{c^{\Gamma_{25}}}{\sqrt{2}} R_2(r) \begin{pmatrix} 2Y_{2,1} \\ 0 \\ 0 \\ Y_{2,2} - Y_{2,-2} \end{pmatrix} + \frac{c^{\Gamma_{12}}}{\sqrt{2}} R_2(r) \begin{pmatrix} 0 \\ -\sqrt{2} Y_{2,0} \\ 0 \\ Y_{2,2} + Y_{2,-2} \end{pmatrix}, \tag{2}$$

which now has a cubic symmetry such as the top of the valence bands in a zincblende crystal. Note that, because of time-reversal symmetry, the states $\Psi_{-3/2}$ and $\Psi_{-1/2}$ are Kramers conjugates of $\Psi_{3/2}$ and $\Psi_{1/2}$, therefore they are not shown here explicitly. The spherical harmonics terms

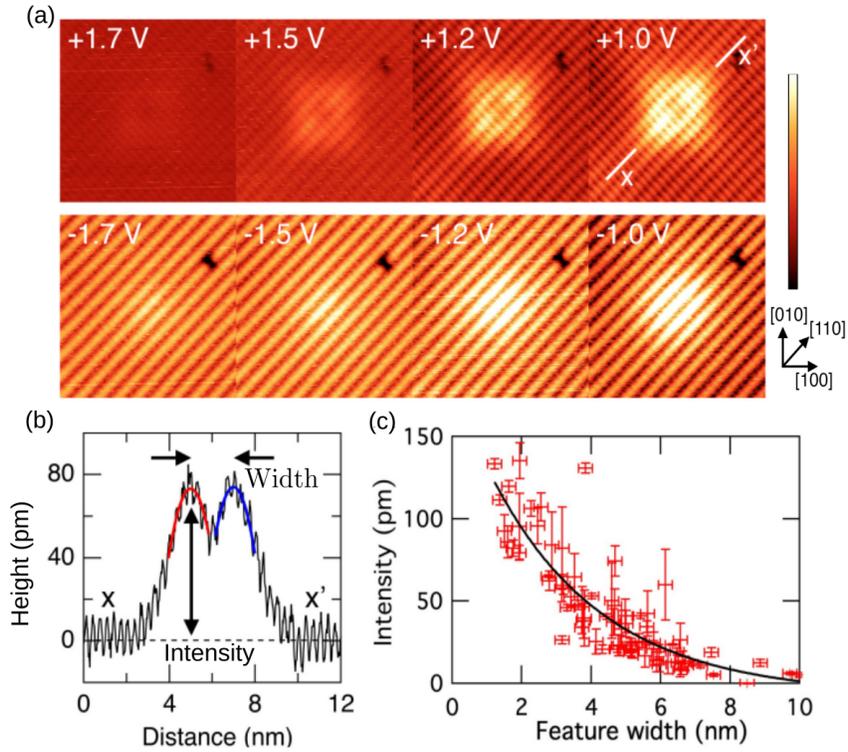

Figure S2: (a) Empty and filled-state images (10×10 nm$^2$) of a single acceptor over a range of imaging bias from ±1.7 to ±1.0 V. Image set point, 20 pA. The imaging set point was 20 pA. The z-range is 150 pm for the empty-state images and 80 pm for the filled-state images. (b) Example line profile over a single feature illustrating how the feature intensity and width are extracted. (c) Plot of relative intensity versus width for 75 acceptor states. The images were recorded with tunneling bias in the range +0.9 to +1.2 V, and with different tips, resulting in a ±10 pm contribution to the uncertainty of the intensities.

$Y_{l,m} \equiv Y_{l,m}(r,\theta,\phi)$ that go along with the coefficients $c^{\Gamma_1}$, $c^{\Gamma_{25}}$ and $c^{\Gamma_{12}}$ transform like $\Gamma_1$(s-like), $\Gamma_{12}$ ($d_{x^2-y^2}$ - like) and $\Gamma_{25}$ ($d_{xy}$ - like), respectively. In analogy what is done in [6], we consider $c^{\Gamma_1} = 1$ and the coefficients $c^{\Gamma_{25}}$ and $c^{\Gamma_{12}}$ are parameterized by $\eta$ (with $0 \leq \eta \leq 1$) as

$$c^{\Gamma_{25}} = \frac{1}{\sqrt{3+2\eta^2}}, \quad c^{\Gamma_{12}} = \frac{\eta}{\sqrt{3+2\eta^2}}. \tag{3}$$

For $\eta = 0$ ($\eta = 1$) the solution above has a cubic (spherical) symmetry. The radial functions in Eqs.(1) and (2) have the following form

$$R_0(r) = \frac{C_0}{r}\left[\beta\, e^{-qr\sqrt{\beta}} + e^{-qr}\right], \tag{4}$$

$$R_2(r) = \frac{C_0}{r}\left[-\beta\, e^{-qr\sqrt{\beta}}\left(1 + \frac{3}{qr\sqrt{\beta}} + \frac{3}{q^2 r^2 \beta}\right) + e^{-qr}\left(1 + \frac{3}{qr} + \frac{3}{q^2 r^2}\right)\right]. \tag{5}$$

with the constants $C_0$, $\beta$ and $q$ determined by

$$C_0 = \sqrt{\frac{q}{\beta\sqrt{\beta}+1}}, \quad \beta = \frac{m_{lh}}{m_{hh}}, \quad q = \sqrt{\frac{2\,m_{hh}E_0}{\hbar^2}}, \tag{6}$$

where $m_{lh}$ and $m_{hh}$ are the light and heavy hole masses determined within the spherical approximation for the Luttinger parameters, $m_0$ is the electron mass and $E_0$ is the binding energy.

The average solution that represents the acceptor states measured with STM is

$$\left|\Psi_{\text{avg}}\right|^2 = \frac{1}{2}\left(\left|\Psi_{1/2}\right|^2 + \left|\Psi_{3/2}\right|^2\right), \tag{7}$$

and using Eqs.(1) and (2), we find

$$\left|\Psi_{\text{avg}}(r,\theta,\phi)\right|^2 = \left|R_0(r)\right|^2\left|Y_{0,0}(\theta,\phi)\right|^2 + \left|R_2(r)\right|^2\Big[\frac{\eta^2}{(3+2\eta^2)}\left|Y_{2,0}(\theta,\phi)\right|^2 + \frac{2}{(3+2\eta^2)}\left|Y_{2,1}(\theta,\phi)\right|^2$$
$$+ \left(\frac{(\eta^2+1)+(\eta^2-1)\cos(4\phi)}{3+2\eta^2}\right)\left|Y_{2,2}(\theta,\phi)\right|^2\Big]. \tag{8}$$

Observe that, for $\eta = 1$ the Eq.(8) reduces to spherical solution given by

$$\left|\Psi_{\text{avg}}(r)\right|^2 = \frac{1}{4\pi}\left(\left|R_0(r)\right|^2 + \left|R_2(r)\right|^2\right) \tag{9}$$

In contrast, for $\eta = 0$, Eq.(8) has an angular dependency that multiplies the $R_2(r)$ term

$$\left|\Psi_{\text{avg}}(r,\theta,\phi)\right|^2 = \frac{1}{4\pi}\left|R_0(r)\right|^2 + \frac{5}{4\pi}\left|R_2(r)\right|^2\left[\cos^2(\theta)\sin^2(\theta) + \sin^2(\phi)\cos^2(\phi)\sin^4(\theta)\right]. \tag{10}$$

Graphically, the solutions for the spherical $\eta = 1$ and cubic $\eta = 0$ given by Eqs.(9) and (10), respectively, are shown in Supplementary Fig.S3.

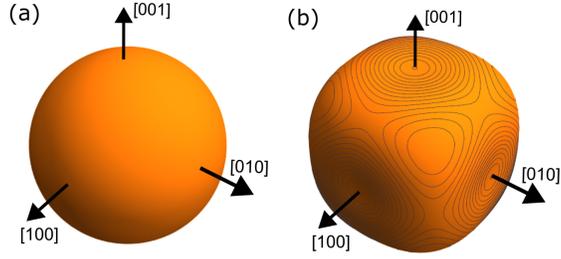

Figure S3: a) and b) show $|\Psi_{\text{avg}}|^2\, e^{qr} = 5\times 10^{-4}\,\text{cm}^{-1}$ for $\eta = 1$ and $\eta = 0$, respectively. Note that when $\eta \to 0$ the spherical symmetry is no longer present.

Table S2: Parameters used for plots in Supplementary Figs. S3 and S4. $m_{lh}$ and $m_{hh}$ are calculated considering the spherical approximation for the Luttinger parameters, and $E_0$ is the expected binding energy for the ground-state acceptor in silicon (Si).

| Material | $\gamma_1$ | $\gamma_2$ | $\gamma_3$ | $m_{lh}/m_0$ | $m_{hh}/m_0$ | $\beta$ | $E_0$ (meV) |
|---|---|---|---|---|---|---|---|
| Si | 4.22 | 0.39 | 1.44 | 0.159 | 0.458 | 0.348 | 30 |

The experimental images show strong electronic contrasts with a square ring-like feature at planes along the [001] direction, through the height from the tip to the surface. In analogy to this, we use the following relation

$$\text{Height}(r, \theta, \varphi) = \frac{1}{2\kappa} \ln\left( \frac{|\Psi_{\text{avg}}(r, \theta, \varphi)|^2}{|\Psi_{\text{avg}}(r_0, \theta_0, \varphi_0)|^2} \right). \quad (11)$$

Here $(r_0, \theta_0, \varphi_0)$ denotes the point at which Height $= 0$ and $\kappa$ is the tunnel decay length constant, which usually in STM measurements is $\sim 1\,\text{Å}^{-1}$.

### S2.2. Supplementary EFMA results

Supplementary Fig. S4 presents the Height$(r, \theta, \varphi)$, Eq.(11), using parameters from Tab. S2. The effective mass result in Fig. S4a shows qualitatively a 'square ring' shape, similar to the STM image in Fig.2 present in the paper. However, comparing the Height profile in Fig. S4b with the profile obtained from STM (Fig.2 from the main text), the width and dip from the calculations are off from the experimental results.

To obtain the results in Fig. S4, the binding energy used was $E_0 = 30$ meV based on the STM results. A similar value is found from the shallow-state solution of the Luttinger-Kohn Hamiltonian for a Coulomb potential [2]. The value for $\beta$ assumes the spherical approximation for the Luttinger parameters, see Tab. S2. However, the actual Si samples have a terminating (001) surface at which the STM measurements are performed; surfaces can cause the appearance of strain [4] which may change the light- and heavy-hole masses $m_{lh}$ and $m_{hh}$, respectively.

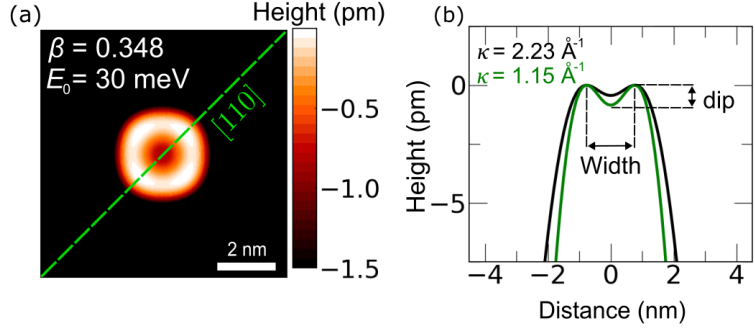

Figure S4: a) shows the height at a (001) plane 5 nm distant from the impurity site, the green line indicates the [110] direction of the profile plot in b), where width = 1.5 nm and dip = 1.0 pm. Tab.S2 contains the parameters for these calculations. a) is qualitatively similar to the STM images, however from the profile line in (b) the 'square ring' feature does not have the same dip and width as observed from the STM profile in Fig.2 from the paper, which has dip $\approx$ 30 pm for a width $\approx$ 2 nm.

Considering the above mentioned problems, we explored our model in order to identify a possible set of optimal parameters for $m_{lh}$ and $m_{hh}$ that could reproduce the dip and width measured. Therefore, the Height was calculated at different planes along the [001] direction considering various values for $\beta$ and $\kappa$, see Figs. S5 - S7.

Fig. S5 shows that the dip at different [001] planes acquires a peak, and its value for a given $\kappa$ depends only on the ratio between light and heavy hole masses $\beta$. As $\beta$ decreases from 0.348 the peak value for the dip increases. If $\beta$ increases the peak value decreases, e.g., fixing $m_{hh} = 0.688$ and increasing $m_{lh}$ from 0.08 in Fig.S6a to 0.319 in Fig.S6d we clearly see the peak value decreasing. Something similar happens if $m_{lh}$ is fixed (e.g., $m_{lh} = 0.16$) and $m_{hh}$ decreases, check Fig.S6b.

Fig. S6 presents a similar behavior for the width. For a given $\kappa$, as $\beta$ decreases from 0.348, the peak value for the width increases. Furthermore, as $\beta$ decreases, the peak value moves closer to the impurity site.

To find an optimal value for $m_{lh}$, $m_{hh}$ and $\kappa$, we need to combine the results from Figs. S6 and Fig. S6, such that for a given width we have its respective dip. Therefore Fig.S7 summarizes our analyses, with dip versus width at different planes. A similar image is shown Fig.2 in the paper, where $m_{lh} = 0.16$ (i.e., its bulk value) and after varying $m_{hh}$ we find that $m_{hh} = 1.261$ for a plane 2.9 nm deeper from the impurity site, produce a square-ring feature with dip= 30 pm and width= 2.0 nm, similar to the experimental values. $m_{lh}$ was fixed based on the idea that the light-hole mass is likely less affected by strain than the heavy-hole mass. For example, in Ref. [3] the Luttinger-Kohn Hamiltonian was solved for heterostructures under strain, and $m_{lh}$ was found to be less susceptible to strain than $m_{hh}$.

7

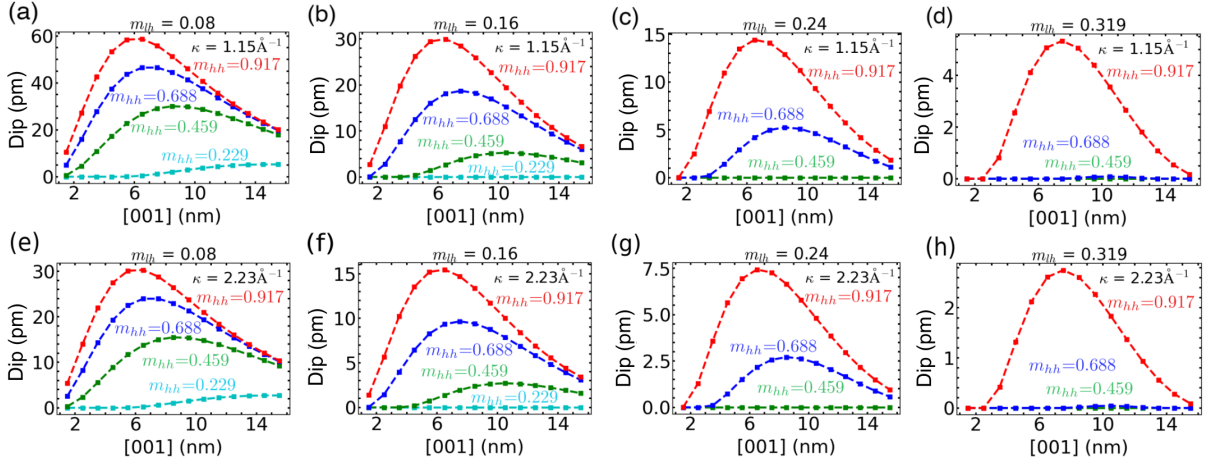

Figure S5: dip versus the plane distance [001] for different values of $m_{lh}$ and $m_{hh}$ (both normalized by $m_0$) using $\kappa = 1.15$ Å$^{-1}$ in a), b), c) and $\kappa = 2.23$ Å$^{-1}$ in d), e) and f).

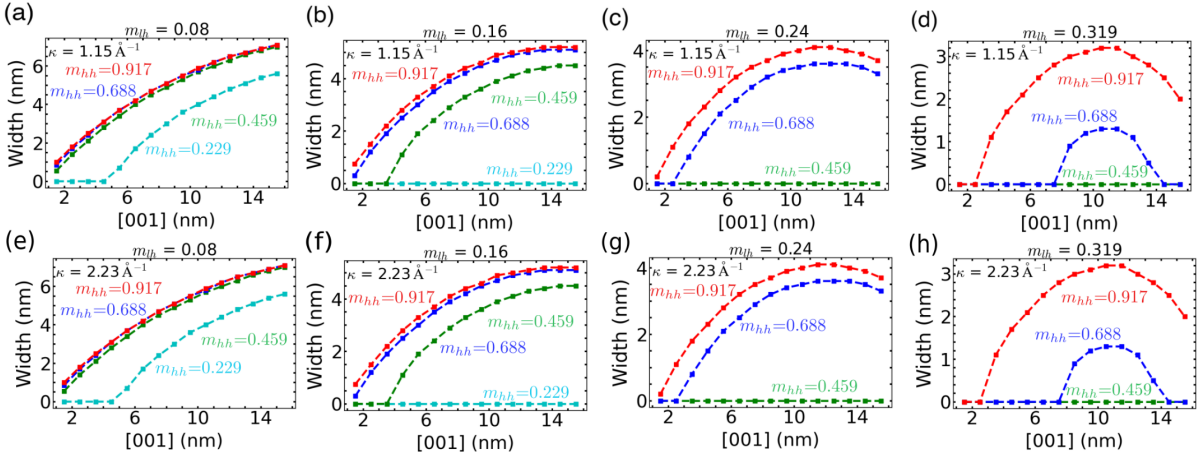

Figure S6: Width versus the plane distance [001] for different values of $m_{lh}$ and $m_{hh}$ (both normalized by $m_0$) using $\kappa = 1.15$ Å$^{-1}$ in a), b), c) and $\kappa = 2.23$ Å$^{-1}$ in d), e) and f).

### S2.3. Tight-Binding Green's Function

For states near a band edge, a decomposition in terms of Bloch states often can be shown to have a simple dependence on a single band and a small wavenumber $\vec{k}$. As states move deeper into the band gap the range of $\vec{k}$ and the number of bands required to effectively model the state increases. In such cases, it becomes necessary to utilize multiband electronic models that are accurate over a wide range of crystal momenta, and ideally the entire Brilliuon zone.

To study localized point defects, we solve the Dyson equation,

$$\hat{G}(\omega) = (1 - \hat{V}'\hat{g}(\omega))^{-1}\hat{g}(\omega) \qquad (12)$$

8

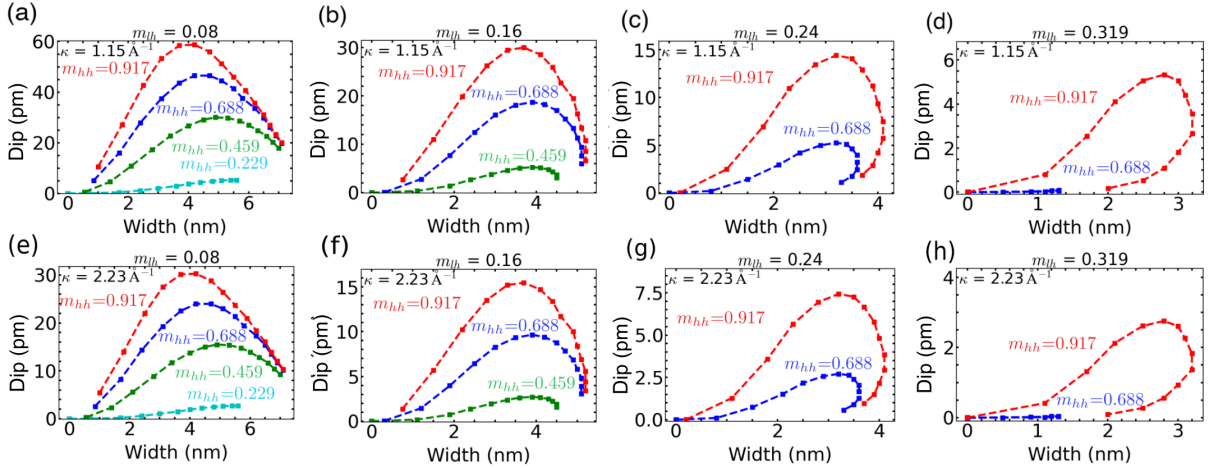

Figure S7: dip versus width for different values of $m_{lh}$ and $m_{hh}$ (both normalized by $m_0$) using $\kappa = 1.15$ Å$^{-1}$ in a), b), c) and $\kappa = 2.23$ Å$^{-1}$ in d), e) and f)..

for the inhomogeneous Green's functions, which correspond to the amplitudes with which bulk Bloch functions scatter off of the localized defect potential, $V'$.

The real-space bulk Green's functions, $g(\omega)$, are computed by taking the numerical inverse Fourier transform of the resolvent of the reciprocal-space Hamiltonian,

$$g(\mathbf{r},\mathbf{r}';z) = \int_{BZ}[z - \hat{H}(\mathbf{k})]^{-1}e^{i(\mathbf{r}-\mathbf{r}')\cdot\mathbf{k}}d\mathbf{k} \tag{13}$$

where $z = \omega + i\delta$. The small imaginary contribution from $\delta$ is necessary to avoid the singularity that arises when $\omega = E(\mathbf{k})$, where $E(\mathbf{k})$ is the energy dispersion found by diagonalizing $H(\mathbf{k})$. The $\delta$ value corresponds to a minimum line-width that limits the maximum resolution obtained for spectral features in the calculation.

For a defect potential that extends over a finite range of atoms in the crystal, the Dyson equation can be solved exactly by separating the problem into two parts, namely, the so-called "near field" and "far field". The near field is defined to contain all atoms where the defect potential has a non-zero value, and the far field is defined to encompass the remainder of atoms in the calculation (where the defect potential is zero). In this basis, our operators can be expressed in block form as,

$$\hat{G} = \begin{pmatrix} \hat{G}_{nn} & \hat{G}_{nf} \\ \hat{G}_{fn} & \hat{G}_{ff} \end{pmatrix} \qquad \hat{g} = \begin{pmatrix} \hat{g}_{nn} & \hat{g}_{nf} \\ \hat{g}_{fn} & \hat{g}_{ff} \end{pmatrix} \qquad \hat{V} = \begin{pmatrix} \hat{V}'_{nn} & 0 \\ 0 & 0 \end{pmatrix} \tag{14}$$

Plugging in the block form in Equation 14 into Equation 12, we end up with an exactly solvable set

9

of equations for the inhomogeneous Green's functions,

$$\hat{G} = \begin{pmatrix} \hat{G}_{nn} & \hat{G}_{nf} \\ \hat{G}_{fn} & \hat{G}_{ff} \end{pmatrix} = \begin{pmatrix} \hat{M}_{nn}\hat{g}_{nn} & \hat{M}_{nn}\hat{g}_{nf} \\ \hat{g}_{fn}\hat{M}_{nn} & \hat{g}_{ff} + \hat{g}_{fn}\hat{V}'_{nn}\hat{M}_{nn}\hat{g}_{nf} \end{pmatrix} \quad (15)$$

where

$$\hat{M}_{n,n} = (1 - \hat{g}_{n,n}\hat{V}'_{n,n})^{-1} \quad (16)$$

We can see from Equation 15 that the computation of the inhomogeneous Green's functions depends only on the local propagator ($g_{ii}$), the propagator from the local site to the defect impurity ($g_{i,n}$), and the description of the local impurity in the near field defined by $M_{nn}$. As there is no coupling of terms in the far field, this method is free from artifacts introduced by methods relying on exact diagonalization, such as finite volume effects and descriptions of surface terminations.

To generate the simulated topographical scans, we first need to extract the local density of states, $\eta$, from the inhomogeneous Green's functions. $\eta$ is related to the imaginary part of the Green's function via the relation

$$\eta(\mathbf{r};\omega) = \frac{-1}{\pi} Im[Tr[\hat{G}(\mathbf{r},\mathbf{r};\omega)]]. \quad (17)$$

Simulated topographical scans of defects are generated by computing $\eta$ for every atomic site (i.e., $\mathbf{r}$) in the field of view at a given plane height, and then convolving with a spherical Gaussian of FWHM of one-quarter the nearest-neighbor bond distance ($d_{NN} = 2.35$Å).

### S2.4. Supplementary tight binding Green's function results

We performed an additional calculation assuming a split-vacancy model with a bound state at 30 meV. The resulting wave function has $S_6$ site symmetry and can be seen in Fig.S8 to strongly deviate from the experimental square-ring spatial structure. Our model of a split vacancy configuration for the defect acceptor state is thus inconsistent with available experimental data.

## References

[1] AVERKIEV, N., AND IL'INSKII, S. Y. Spin ordering of carriers localized at two deep centers in cubic semiconductors. *Physics of the Solid State 36*, 2 (1994), 278–283.

[2] BALDERESCHI, A., AND LIPARI, N. O. Spherical model of shallow acceptor states in semiconductors. *Physical Review B 8*, 6 (1973), 2697.

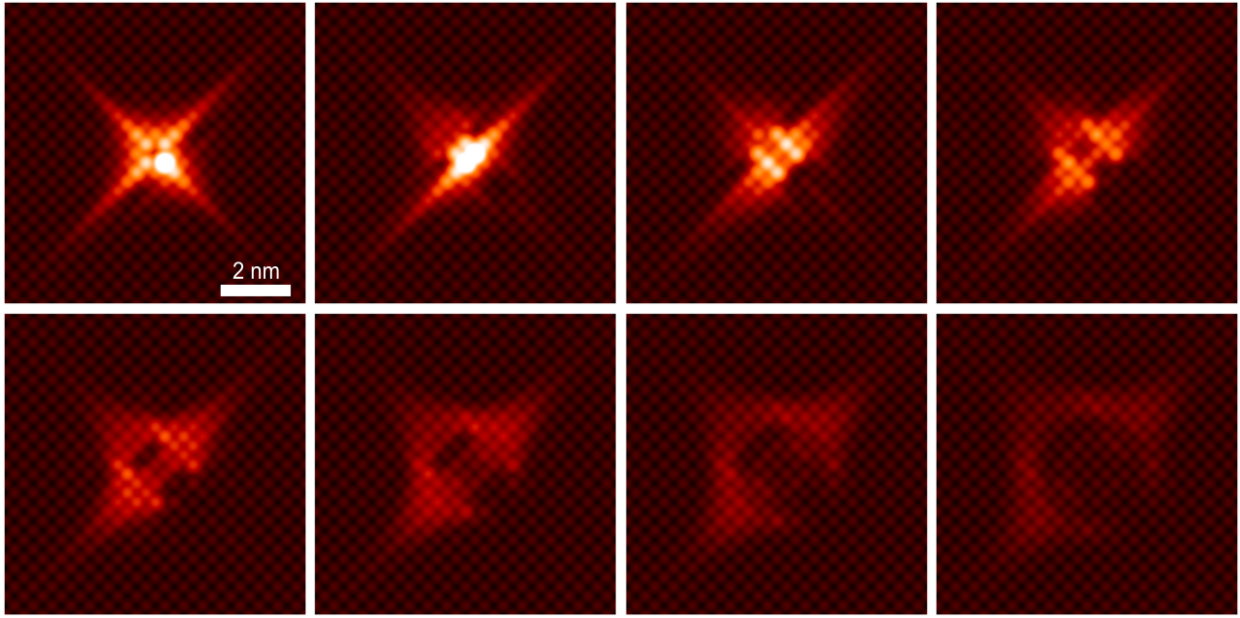

Figure S8: Series showing the acceptor spatial structure as a function of layer depth for a split-vacancy defect. The spatial structure is skewed towards the upper-left quadrant, reflecting the strong $S_6$ site symmetry. The layer spacing along (001) is 1.357Å.

[3] Chuang, S. L. Efficient band-structure calculations of strained quantum wells. *Phys. Rev. B 43* (Apr 1991), 9649–9661.

[4] Çelebi, C., Garleff, J. K., Silov, A. Y., Yakunin, A. M., Koenraad, P. M., Van Roy, W., Tang, J.-M., and Flatté, M. E. Surface induced asymmetry of acceptor wave functions. *Phys. Rev. Lett. 104* (Feb 2010), 086404.

[5] Siegl, M. *Atomic-scale investigation of point defect interactions in semiconductors*. PhD thesis, UCL (University College London), 2018.

[6] Yakunin, A., Silov, A. Y., Koenraad, P. M., Wolter, J., Van Roy, W., De Boeck, J., Tang, J.-M., and Flatté, M. E. Spatial Structure of an Individual Mn Acceptor in GaAs. *Phys. Rev. Lett. 92*, 21 (may 2004), 216806.

[7] Ziegler, J. F., Ziegler, M. D., and Biersack, J. P. SRIM - The stopping and range of ions in matter (2010). *Nuclear Instruments and Methods in Physics Research B 268* (June 2010), 1818–1823.


\end{document}